\documentclass[12pt]{article}
\usepackage{a4wide}
\usepackage{amssymb}
\usepackage{amsmath}
\usepackage{mathrsfs}
\usepackage{graphicx}
\begin{document}
{\renewcommand{\thefootnote}{\fnsymbol{footnote}}
\medskip
\begin{center}
{\LARGE  On anomaly freedom in spherically symmetric lattice loop quantization}\\
\vspace{1.5em}
Mikhail Kagan\footnote{e-mail address: {\tt mak411@psu.edu}}
\\
\vspace{0.5em}
The Pennsylvania State University, Abington College\\
\vspace{1.5em}
\end{center}
}

\setcounter{footnote}{0}
\newcommand{\Lam}{\Lambda}
\newcommand{\vt}{\vartheta}
\newcommand{\be}{\begin{equation}}
\newcommand{\ee}{\end{equation}} 
\newcommand{\bq}{\begin{eqnarray}} 
\newcommand{\eq}{\end{eqnarray}}
\newcommand{\f}{\frac}
\newcommand{\vp}{\varphi}
\newcommand{\abs}[1]{\lvert#1\rvert}

\newcommand{\case}[2]{{\textstyle \frac{#1}{#2}}}
\newcommand{\lP}{\ell_{\mathrm P}}

\newcommand{\md}{{\mathrm{d}}}
\newcommand{\Kern}{\mathop{\mathrm{ker}}}
\newcommand{\tr}{\mathop{\mathrm{tr}}}
\newcommand{\sgn}{\mathop{\mathrm{sgn}}\nolimits}

\newcommand*{\R}{{\mathbb R}}
\newcommand*{\N}{{\mathbb N}}
\newcommand*{\Z}{{\mathbb Z}}
\newcommand*{\Q}{{\mathbb Q}}
\newcommand*{\C}{{\mathbb C}}
\def\v{{\rm{\bf v}}}
\def\I{{\rm{\bf I}}}
\def\J{{\rm{\bf J}}}
\def\K{{\rm{\bf K}}}
\def\H{{\cal H}}
\def\D{{\cal D}}
\def\G{{\cal G}}
\newcommand{\p}{{\partial}}

\begin{abstract}
Anomaly freedom has been one of the most important issues in canonical quantization of gravity. In a physically meaningful (anomaly free) theory,  the constraint operators must be first class, and their commutator algebra is expected to resemble the corresponding classical Poisson-bracket algebra. In this paper, we review the ``constructive\rq{}\rq{} approach to obtaining a consistent set of constraints: start with a Hamiltonian constraint and generate the corresponding diffeomorphism constraint as a commutator of two Hamiltonians. Closure of the constraint operator algebra then requires that the diffeomorphism operator obtained in this way weakly commutes  with another Hamiltonian constraint operator. The same procedure can be used to check the consistency of some proposed quantization schemes that present a candidate Hamiltonian constraint, which we do for a spherically symmetric model.
\end{abstract}

\section*{Introduction}
The first models for canonical quantization of gravity go back to the ADM formalism \cite{ADM, ADMRe}. The spacetime metric is decomposed into a spatial metric, lapse function and shift vector, which leads to going from the manifestly covariant Einstein-Hilber action to a system of constraints of canonical variables and a Poisson bracket between those variables. The canonical equations of motion are generated using the constraints and Poisson brackets. Keeping the lapse function and shift vector arbitrary allows to ensure general covariance at the classical level and, as a result, consistency and gauge invariance of the resulting equations of motion. 

The classical Hamiltonian and diffeomorphism constraints form a closed (first class)  algebra (see Eqs. (\ref{DD})-(\ref{HH})). When quantizing the classical system of constraints using Dirac\rq{}s procedure \cite{DirQuant}, one needs to make sure that the commutator algebra of the constraint operators remains first class. In this case, the quantum system is said to be anomaly free. The same requirements apply to effective theories, as outlined in \cite{ConstraintAlgebra} and \cite{ScalarGaugeInv}.

In the framework of loop quantum gravity, it has been shown that the spectra of the geometrical operators, such as area and volume, are fundamentally discrete.  Therefore, 
it appears natural to consider quantization of the physically interesting constraint operators on a discrete structure, a spin network \cite{Rovelli_Smolin}.  A suitable inner product for quantum states in this framework was suggested by Ashtekar and Lewandowski in \cite{AL-measure} and the relevant operators were shown to be self-adjoint. Recently anomaly-free quantization of the loop quantum gravity constraints in three spacetime dimensions has been considered in \cite{ConstAlg_HLT-1}-\cite{AF_QDyn}.

In this paper, we consider spin networks of regular structure, i.e. lattices that are motivated by the symmetry of the model. The two simple examples include cubic and spherically symmetric lattices. For the quantization purposes, we take on a {\em constructive} approach: we originally construct only the Hamiltonian constraint motivated by \cite{QSDI, SphericalQG_Ham, JR}. In the spirit of the classical constraint algebra (see Eq.(\ref{HH})), we then {\em define} the Diffeomorphism constraint operator using the commutator of two Hamiltonian constraints. The system of constaints constructed in this way will be an adequate quantization of the classical system if the Diffeomorphism constraint (i) weakly commutes with the Hamiltonian, and (ii) has the right classical limit. The same procedure can be used to check the consistency of some proposed quantization schemes that present a candidate Hamiltonian constraint. We follow the outlined procedure for the construction motivated by the one in \cite{SphericalQG_Ham}, although of somewhat more general form. We find that in the considered framework, the closure of the quantum constraint algebra is problematic.

The paper is organized as follows. In Section \ref{Sec:Class_algebra}, we review the classical hypersurface deformation algebra for the full theory and its restriction to the spherically symmetric sector. We then proceed to describing the lattice quantization procedure in Section \ref{Sec:Lattice_quantization} by, first, using cubic lattice as an example and a visual guidance for the spherically symmetric case. The descrption of the basic states and operators is followed by the construction of the Hamiltonian constraint and derivation of Diffeomorphism constraint, which is shown to have he right classical limit. In Section \ref{Sec:Closure}, we investigate whether the two constraints can form a  first class algebra and conclude with a discussion in Section \ref{Sec:Discussion}.

\section{Classical constraint algebra}\label{Sec:Class_algebra}
In this section we review the classical formulation of canonical loop quantum gravity. Starting with the basic variables, constraints and their algebra of the full theory, we proceed to the spherically symmetric sector.
\subsection{Full theory}
Canonical (ADM) approach to gravity \cite{ADM,ADMRe} is  based on splitting the space-time metric into spatial metric $q_{ab}$, lapse function $N$ and shift vector $N^a$ ($a=1,2,3$)
\begin{equation} \label{ds}
 {\rm d}s^2= -N^2{\rm d}t^2+ q_{ab} ({\rm d}x^a+N^a{\rm d}t) ({\rm
   d}x^b+N^b{\rm d}t).
\end{equation}
In LQG one further rewrites the (inverse) spatial metric in terms of the densitized triad 
\[
 q^{ab} = \frac{E_i^a E_i^b}{\det E},
\] 
which becomes a basic canonical variable. Its conjugate is the Ashtekar connection
\be\label{AshCon}
A_a^i = \Gamma_a^i +\gamma K_a^i,
\ee
where $\Gamma_a^i$ is the spin connection compatible with the densitized triad and $K_a^i$ is the extrinsic curvature. Performing the space-time splitting of the Einstein-Hilbert action, one obtains
\be
S=\int {\rm d}t \left[\frac{1}{8\pi G\gamma}\int{{\rm d}x^3 E^a_i\dot A_a^i}-\int{{\rm d}x^3 \left(N\H+N^a\D+\lambda^i \G_i\right)}\right],
\ee
where $G$ is the gravitational constant and $\gamma$ is the Barbero-Immirzi parameter. From the first term we gather that the Poisson bracket between the densitized triad and Ashtekar connection is given by
\be\label{PB_AE}
\left\{A_a^i(x),E_j^b(y)\right\}=8\pi G \gamma \delta_a^b\delta_j^i\delta(x,y).
\ee
The second integral contains Lagrange multipliers and three constraints that generate gauge transformations. The Hamiltonian constraint density $\H$ gives rise to coordinate time evolution; the diffeomorphism constraint $\D$ generates spatial diffeomorphism transformations, and the Gauss constraint $\G$ is responsible for internal rotations of the triad that do not affect the spatial metric. The explicit expressions for the three constraints are as follows.
\bq\label{Ham_Full}
H_{\rm grav}[N]&=&\frac{1}{16 \pi G} \int {\rm d}x^3 N\frac{E_i^a E_j^b}{\sqrt{\det E}}\left[\epsilon^{ij}{}_kF_{ab}^k-2(1+\gamma^2)K^i_{[a}K^j_{b]} \right],\\
D_{\rm grav}[N^a]&=&\frac{1}{8 \pi G \gamma}\int {\rm d}x^3 N^a \left[\left(\p_a A_b^j - \p_b A_a^j\right)E_j^b-A_a^j\p_b E_j^b\right], \label{Diff_Full}\\
G_{\rm grav} [\lambda^i]&=& \frac{1}{8 \pi G \gamma}\int {\rm d}x^3\left(\p_a E_i^a +\epsilon_{ij}{}^k A_a^j E_k^a\right).
\eq
Here $F_{ab}^k = 2\p_{[a}A_{b]}^k+\epsilon_{ij}{}^k A_a^i A_b^j$ is the curvature of the Ashtekar connection. These constraints are first class and their Poisson brackets are weakly zero. The Gauss constraint is often solved explcitly to eliminate the spin connection in terms of the densitized triad. The remaining two constraints satisfy the following relations
\begin{eqnarray}
 \{D[N^a],D[M^b]\} &=& -D[{\cal
   L}_{M^b}N^a]\,,\label{DD}\\
\{H[N],D[N^a]\} &=& -H[{\cal L}_{N^a}N]\label{HD}\,,\\
\{H[N],H[M]\} &=& D[N{\p^a}M-M{\p^a}N]\,. \label{HH}
\end{eqnarray}
In order to obtain the contravariant shift vector in the last diffeomorphism constraint, one needs to use either spatial metric or densitized triad to raise the indices of the covariant derivatives.
\subsection{Spherical symmetry}
Below we review the basic variables and the corresponding constraints in the spherically symmetric setting. Our construction and notation will largely follow that in Refs. \cite{SphericalQG_Ham, JR}.
\subsubsection{Basic variables}
Let $x$ and $\varphi$ represent the radial and angular coordinates respectively. The basic canonical variables are the corresponding components of the triad and extrinsic curvature: $(E^\varphi,K_\varphi)$ and $(E^x,K_x)$. In terms of these triad variables, the spatial (spherically symmetric) metric is given by
\be
\md q^2 = \frac{E_\varphi^2}{\left|E_x \right|}\md x^2 + \left|E^x \right|\md \Omega^2.
\ee
Solving the Gauss constraint yields an explicit expression for the spin connection in terms of the triads (and their derivatives). The angular component of the spin-connection is
\be\label{Sph_Spin_Conn}
\Gamma_\varphi=-\frac{E^{x\prime}}{2E^\varphi},
\ee
where the prime denotes the $x-$derivative. $\Gamma_\varphi$ is gauge invariant, whereas its radial component $\Gamma_x$ is pure gauge. The canonical variables satisfy the following Poisson brackets
\be\label{Sph_PB}
\case{1}{2}\left\{K_x(x),E^x(y)\right\}=\left\{K_\varphi(x), E^\varphi(y)\right\}=8\pi G\delta(x,y).
\ee
\subsubsection{Constraints}
The spherically symmetric version of the Hamiltonian and diffeomorphism constraints in Eqs. (\ref{Ham_Full}) and (\ref{Diff_Full}) is given by 
%
\be
H_{\rm G}[N] =-\f{1}{2G}\int{\md x N |E^x|^{-1/2}}\left[K_\varphi^2E^\varphi + 2 K_\varphi K_x E^x+\left(1-\Gamma^2_\varphi\right)E^\varphi+2\Gamma^\prime_\varphi E^x\right]
\ee
and
\be\label{ClassDiff}
D_{\rm G}[N^x]=\f{1}{2G}\int{\md x N^x \left[ 2E^\varphi K_\varphi^\prime -K_xE^{x\prime}\right]}
\ee
respectively. Note that a term proportional to the Gauss constraint was removed from the diffeomorphism constraint. These constraints are first class and satisfy the following Poisson brackets
\begin{eqnarray}
 \{D[N^x],D[M^x]\} &=& -D[{M^x}N^{x\prime}-N^xM^{x\prime}]\,,\label{Sph_DD}\\
\{H[N],D[N^x]\} &=& -H[N^xN^\prime]\label{Sph_HD}\,,\\
\{H[N],H[M]\} &=& D[{|E_x|}/{(E^\varphi)^2}\left(NM^\prime-MN^\prime\right)]\, , \label{Sph_HH}
\end{eqnarray}
where the triad components are explicitly present on the righthand side of the last expression. In the next section we attempt to quantize the above system of constraints on a lattice, such that the quantum counterparts of the constraints satisfy analogous commutator algebra.

\section{Lattice quantization}\label{Sec:Lattice_quantization}
The quantization procedure requires promoting the  canonical variables $E^\varphi,E^x, K_\varphi, K_x$ into quantum operators and switching from Poisson brackets to commutators. In the loop quantization, the basic  operators correspond to the smeared versions of the canonical variables, namely fluxes of densitized triads and holonomies of the extrinsic curvature. It is easier to visualize the geometrical structure of the lattice loop quantization in a three-dimensional case. We do so in Section \ref{Sec:Cubic_Lattice} by considering, for simplicity, a cubic lattice. In Section \ref{Sec:Spherical_Lattice}, we then repeat a similar construction for the spherically symmetric model.
\subsection{Cubic Lattice}\label{Sec:Cubic_Lattice}
For illustration purposes, consider the Euclidean part of the Hamiltonian constraint in Eq. (\ref{Ham_Full}). When quantizing, the curvature factor $F_{ab}^k$ is to be replaced by a holonomy around a rectangular loop spanned by the directions $a$ and $b$. At the same time, the triad prefactor can be rewritten using Thiemann\rq{}s trick \cite{QSDI}
\begin{equation}
 \left\{A_c^k,\int{\rm d}^3x\, \sqrt{|\det E^a_j|}\right\}\propto
 \epsilon^{ijk}\epsilon_{abc} \frac{E^a_i E^b_j}{\sqrt{|\det E^d_l|}}\,.
\end{equation}
Here the triad factor is expressed in terms of the Poisson bracket of the volume, whose loop quantum operator is well defined, and an Ashtekar connection, whose operator is not. The proper quantization requires one to use a holonomy along an edge that is perpendicular to the loop used for the curvature holonomy
\[
\frac{E E}{\sqrt{\det E}} \propto \left\{ A, V\right\}\rightarrow \hat h\!\left[\hat{h}^{-1}, \hat V\right],
\]
where the edge holonomy and its inverse are taken along the same edge.
\begin{figure}[h!] 
\centerline{\includegraphics[width=8cm, keepaspectratio]{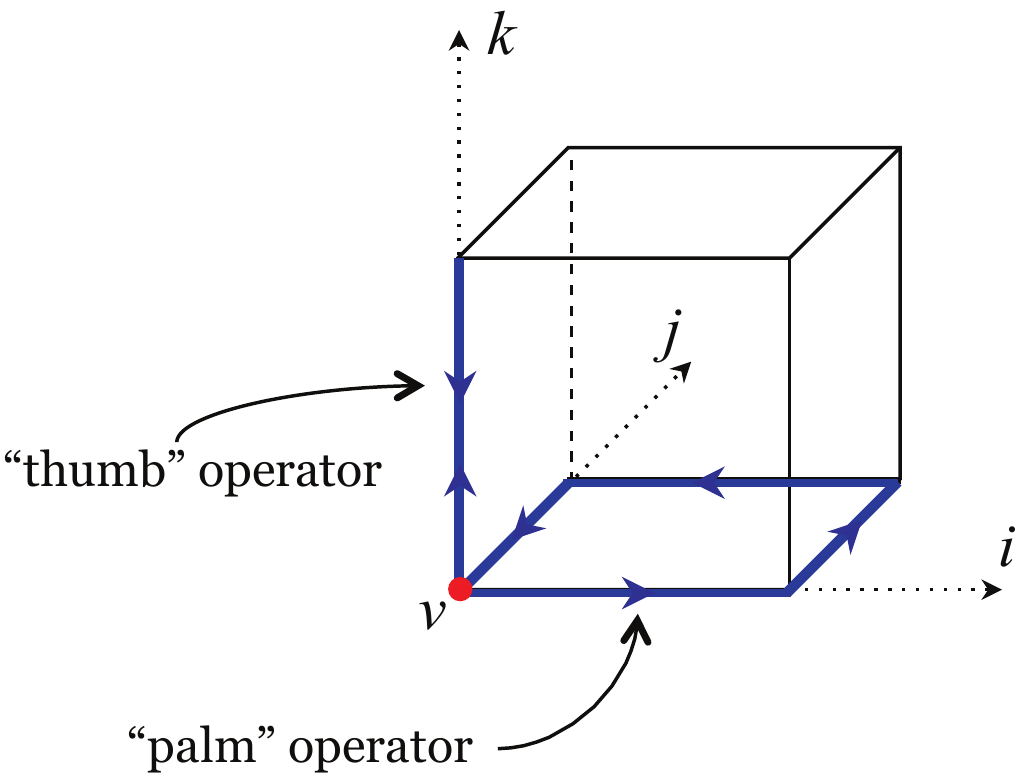}} \caption{One representative \lq\lq{}hand\rq\rq{} operator based at vertex $v$. It consists of the holonomy operator around the bottom face   (\lq\lq{}palm\rq\rq{} operator) and the doubly traced edge along the $k$-direction (\lq\lq{}thumb\rq\rq{} operator).}\label{Fig:Cube}
\end{figure}
Combining the curvature and triad pieces together yields a typical {\em vertex} contribution to the Hamiltonian constraint operator
\be
{\epsilon^{ijk}\tr\left(\hat h_{ij}\hat h_k\left[\hat{h}^{-1}_k, \hat V\right]\right)}.
\ee
The geometrical structure of this operator is depicted in Fig. \ref{Fig:Cube} and resembles a ``hand\rq{}\rq{} with a horizontal ``palm\rq{}\rq{} and vertical \lq\lq{}thumb\rq\rq{} that correspond to the loop holonomy $\hat h_{ij}$ and the doubly traced edge holonomy $\hat h_k$ and its inverse respectively. In order to obtain the total Hamiltonian constraint, we need to symmetrize over all possible orientations of the hand operator (48 combinations for each vertex) and sum over all the vertices in the relevant region. In that sum, each vertex contribution $\hat{H}_v$ comes with a lapse ``function\rq\rq{} $N(v)$ \cite{QuantCorrPert}
\begin{eqnarray} \label{vertex_Ham}
 \hat{H}_v &=& \frac{1}{16\pi G}\frac{2i}{8\pi\gamma
 G\hbar}\frac{N(v)}{8}
\sum_{IJK}\sum_{\sigma_I\in\{\pm1\}} \sigma_1\sigma_2\sigma_3
\epsilon^{IJK}\\
&\times&\tr\left(h_{v,\sigma_II}(A)
h_{v+\sigma_I\I,\sigma_JJ}(A) h_{v+\sigma_J\J,\sigma_II}^{-1}(A)
\right. \nonumber\left.\times h_{v,\sigma_JJ}^{-1}(A) h_{v,\sigma_KK}(A)
[h_{v,\sigma_KK}^{-1}(A),\hat{V}]\right),
\end{eqnarray}
where $h_{v,I}$ ($h_{v,J}$ or $h_{v,K}$) denotes the holonomy along the edge that starts at vertex $v$ and goes in the direction $I$ ($J$ or $K$), whereas the $\sigma$\rq{}s ($\pm 1$) encode the direction in which the corresponding edge is traversed by the holonomy. In the spherically symmetric model, the total Hamiltonian constraint (including matter) has a form very similar to the Euclidean part of the Hamiltonian of the full theory. The ideas outlined in the current section make visualization and bookkeeping of the relevant terms easier even in the spherically symmetric model.
\subsection{Spherically symmetric lattice}\label{Sec:Spherical_Lattice}
In a spherically symmetric context, a lattice is simply a radial line with some vertices (labeled by $v$) and edges (labeled by $e$), while the angular dimensions are suppressed. The triad component $E^x$ is a scalar, whereas $E^\varphi$ is a scalar density of weight one. Thus the corresponding fluxes are given by
\be
F_v[E^x]=E^x(v), \quad F_e[E^\varphi]=\int\limits_e{\!\!E^\varphi \md x}.
\ee
The radial holonomy and angular (point) holomy read 
\be\label{Holonomies}
h_e[K_x]=\exp\left(\case{i}{2}\smallint\limits_e{\!\!\gamma K_x \md x}\right), {\rm and}\quad 
h_v[K_\varphi]=\exp\left(i\delta\gamma K_\varphi(v)\right)
\ee
respectively.
\subsubsection{Basic states and operators}
The Hamiltonian constraint operator is a sum of vertex operators, each of which only acts on one vertex and two neighbouring edges of a lattice state. In order to simplify the notation, we shall only keep the relevant labels and omit the unaffected ones. For an arbitrary vertex $v$, the ket will read
\begin{equation}
|\cdots,\mu_{v-1},k_{v-1},\mu_{v},k_{v},\mu_{v+1},\cdots\rangle=\mbox{
\begin{picture}(200,15)(0,0)
 \put(0,5){\line(1,0){200}}
 \put(50,5){\circle*{5}}
 \put(100,5){\circle*{5}}
 \put(150,5){\circle*{5}}
 \put(50,-3){\makebox(0,0){$\mu_{v-1}$}}
 \put(100,-3){\makebox(0,0){$\mu_v$}}
 \put(150,-3){\makebox(0,0){$\mu_{v+1}$}}
 \put(25,10){\makebox(0,0){$\cdots$}}
 \put(75,12){\makebox(0,0){$k_{v-1}$}}
 \put(125,12){\makebox(0,0){$k_v$}}
 \put(175,10){\makebox(0,0){$\cdots$}}
\end{picture}},
\end{equation}
where the edge label $k_v$ receives the subscript of the vertex to its immediate left. The connection representation of this state is 
\be\label{T_states}
T_{g,k,\mu}\left[K_x, K_\varphi\right] = \prod\limits_{e \in g}\exp\left(\case{i}{2}k_e\smallint_{e}\gamma K_x\md x\right)\prod\limits_{v \in g}\exp\left(i\mu_v\gamma K_\varphi\right).
\ee
In this paper we restrict ourselves to operators that do not create or destroy vertices. Then there are two basic types of operators: a multiplication operator (e.g. the volume operator) and a shift operator (e.g. a holonomy operator). The former does not change the lattice labels and yields the same lattice state mulitplied by a coefficient that, in general, depends on the neighbouring lattice labels:
\begin{equation}\label{MultOp}
\hat f_v|\mu_{v-1},k_{v-1},\mu_{v},k_{v},\mu_{v+1}\rangle=f_v(\mu_{v-1},k_{v-1},\mu_{v},k_{v},\mu_{v+1})\mbox{
\begin{picture}(160,15)(0,0)
 \put(0,5){\line(1,0){160}}
 \put(30,5){\circle*{5}}
 \put(80,5){\circle*{5}}
 \put(130,5){\circle*{5}}
 \put(30,-3){\makebox(0,0){$\mu_{v-1}$}}
 \put(80,-3){\makebox(0,0){$\mu_v$}}
 \put(130,-3){\makebox(0,0){$\mu_{v+1}$}}
 \put(55,12){\makebox(0,0){$k_{v-1}$}}
 \put(105,12){\makebox(0,0){$k_v$}}
\end{picture}}.
\end{equation}
A shift operator does alter some of the lattice labels at and near the vertex it acts on. One of the simplest examples of a shift operator looks like
\begin{equation}\label{ShiftOp}
\hat g_v|\mu_{v-1},k_{v-1},\mu_{v},k_{v},\mu_{v+1}\rangle=\mbox{
\begin{picture}(160,15)(0,0)
 \put(0,5){\line(1,0){160}}
 \put(30,5){\circle*{5}}
 \put(80,5){\circle*{5}}
 \put(130,5){\circle*{5}}
 \put(30,-3){\makebox(0,0){$\mu_{v-1}$}}
 \put(80,-3){\makebox(0,0){$(\mu_v+\delta)$}}
 \put(130,-3){\makebox(0,0){$\mu_{v+1}$}}
 \put(55,12){\makebox(0,0){$k_{v-1}$}}
 \put(105,12){\makebox(0,0){$k_v$}}
\end{picture}}
\end{equation}
and only affects one vertex label $\mu_v$. Constructing products of several shift operators one can obtain an operator that changes several vertex and/or edge labels. In general, a shift and a multiplication operators do not commute. It is easy to see that for the operators in Eqs. (\ref{MultOp}) and (\ref{ShiftOp})
\begin{equation}\label{CR}
\left[\hat f, \hat g\right]=\hat g_v\left(f_v(\mu_{v-1},k_{v-1},\mu_{v}+\delta,k_{v},\mu_{v+1})-f_v(\mu_{v-1},k_{v-1},\mu_{v},k_{v},\mu_{v+1})\right).
\end{equation}

Physically relevant operators are generally composed of combinations of both multiplication and shift operators. In what follows, we shall assume - without loss of generality - that the multiplication part of such operators is on the right of the shifting part. This can always be achieved by utilizing the commutation relations such as the one in Eq. (\ref{CR}). In this paper, we shall primarily be interested in the algebra of operators acting on the lattice states defined above.  For such operators, it will be convenient to use similar diagrammatic notation. For example, the operator from Eq.(\ref{ShiftOp}) can be represented by the following diagram
\be
\hat g_v=\mbox{
\begin{picture}(160,15)(0,0)
 \put(0,5){\line(1,0){160}}
 \put(30,5){\circle*{5}}
 \put(80,5){\circle*{5}}
 \put(130,5){\circle*{5}}
 \put(80,-3){\makebox(0,0){${\bf v}$}}
 \put(80,12){\makebox(0,0){$(+\delta)$}}
\end{picture}
},
\ee
where we suppressed the uneffected vertex labels. Note that in the connection representation, the holonomies defined in Eq. (\ref{Holonomies}) act on the lattice states in (\ref{T_states}) as multiplication operators, which effectively shifts the vertex or edge labels according to 
\be
\hat h_e[K_x]T_{g,k,\mu} = T_{g,k+1,\mu}, \quad \hat h_v[K_\varphi]T_{g,k,\mu} = T_{g,k,\mu+\delta} .
\ee
The physically interesting operators, such as Hamiltonian or diffeomorphism constraints, can be constructed from combinations of the ones discussed above. This is done in the next two sections.
\subsubsection{Hamiltonian constraint operator}\label{sec:HamConst}
On a spherically symmetric lattice, there are only two essentially distinct types of the ``hand\rq\rq{}-operators: with a radial or angular \lq\lq{}thumb\rq\rq{}. Since the thumb-part of the hand-operator is diagonal, the former type of the hand-operator only affects the label of one vertex ($v$). At the same time, a hand-operator with an angular thumb also affects one of the neighbouring vertices ($v+1$ or $v-1$) and the edge connecting that vertex with $v$. This can be seen in the corresponding diagrams Eq. (\ref{Ham_CLR}). In what follows we review the construction explained in  \cite{SphericalQG_Ham}.

Hamiltonian constraint for a spherically symmetric model can be conveniently represented as a discrete sum of the vertex contributions of three types
\be\label{Ham_Def}
\hat H[N]=\sum\limits_v{N_v\left(\hat H^v_C+\hat H^v_R+\hat H^v_L\right)},
\ee
where the central, right and left terms are defined as
\bq\label{H_terms}
\hat H^v_C |\mu_{v-1},k_{v-1},\mu_{v},k_{v},\mu_{v+1}\rangle 
&=& \sum\limits_{\sigma=0,\pm 1}{|\mu_{v-1},k_{v-1},\mu_{v}+2\sigma\delta,k_{v},\mu_{v+1}\rangle}\Phi^v_\sigma \\
\hat H^v_R |\mu_{v-1},k_{v-1},\mu_{v},k_{v},\mu_{v+1}\rangle 
&=& \!\!\sum\limits_{\sigma_{1,2,3}=\pm 1}{\!\!\!\! \sigma_2 \sigma_3 |\mu_{v-1},k_{v-1},\mu_{v}+\sigma_1\delta/2,k_{v}+2\sigma_2,\mu_{v+1}+\sigma_3\delta/2\rangle}R^v \nonumber\\
\hat H^v_L |\mu_{v-1},k_{v-1},\mu_{v},k_{v},\mu_{v+1}\rangle 
&=& \sum\limits_{\sigma_{1,2,3}=\pm 1}{\!\!\!\! \sigma_2 \sigma_3 |\mu_{v-1}+\sigma_3\delta/2,k_{v-1}+2\sigma_2,\mu_{v}+\sigma_1\delta/2,k_{v},\mu_{v+1}\rangle}R^v \nonumber.
\eq
The relevant (parts of the) lattice states from Eq. (\ref{T_states}) are given by
\begin{eqnarray*}
 \langle K_{\vp},K_x|\mu_{v-1},k_{v-1},\mu_v,k_{v},\mu_{v+1}\rangle&:=& \exp(i\mu_-
 \gamma K_{\vp}(v_-))
 \exp\left(\case{1}{2}ik_-\smallint_{v-1}^v\gamma K_x\md x\right)  \exp(i\mu
 \gamma K_{\vp}(v))\\
&& \exp\left(\case{1}{2}ik_+\smallint_{v}^{v+1}
\gamma K_x\md x\right) \exp(i\mu_+ \gamma K_{\vp}(v_+))\,.
\end{eqnarray*} 
Diagrammatically we can rewrite the expressions (\ref{H_terms}) as
\bq\label{Ham_CLR}
\hat H^v_C &=& \sum\limits_{\sigma=0,\pm 1}{\mbox{
\begin{picture}(160,15)(0,0)
 \put(0,5){\line(1,0){160}}
 \put(30,5){\circle*{5}}
 \put(80,5){\circle*{5}}
 \put(130,5){\circle*{5}}
 \put(80,-3){\makebox(0,0){${\bf v}$}}
 \put(80,12){\makebox(0,0){$+2\sigma\delta$}}
\end{picture}
}\Phi^v_\sigma}, \nonumber\\ 
\hat H^v_R &=& \sum\limits_{\sigma_{1,2,3}=\pm 1}{\!\!\!\! \sigma_2 \sigma_3\mbox{
\begin{picture}(240,15)(0,0)
 \put(0,5){\line(1,0){240}}
 \put(45,5){\circle*{5}}
 \put(120,5){\circle*{5}}
 \put(195,5){\circle*{5}}
 \put(120,-3){\makebox(0,0){${\bf v}$}}
 \put(120,15){\makebox(0,0){$+\sigma_1\tfrac{\delta}{2}$}}
 \put(195,15){\makebox(0,0){$+\sigma_3\tfrac{\delta}{2}$}}
 \put(158,-3){\makebox(0,0){$+2\sigma_2$}}
\end{picture}
}R^v}, \\
\hat H^v_L &=& \sum\limits_{\sigma_{1,2,3}=\pm 1}{\!\!\!\! \sigma_2 \sigma_3\mbox{
\begin{picture}(240,15)(0,0)
 \put(0,5){\line(1,0){240}}
 \put(45,5){\circle*{5}}
 \put(120,5){\circle*{5}}
 \put(195,5){\circle*{5}}
\put(45,15){\makebox(0,0){$+\sigma_3\tfrac{\delta}{2}$}}
  \put(82,-3){\makebox(0,0){$+2\sigma_2$}}
 \put(120,-3){\makebox(0,0){${\bf v}$}}
 \put(120,15){\makebox(0,0){$+\sigma_1\tfrac{\delta}{2}$}}
\end{picture}
}R^v},\nonumber
\eq
where the shifting part of each operator is represented by a quantity above the lattice for the corresponding vertex label and by a quantity below the lattice for the corresponding edge label. Here $\Phi_\sigma^v$ and $R^v$ are functions of the labels at the corresponding node ($\mu_v$) and its adjacent edges ($k_{v-1}$ and $k_v$). Note that in \cite{SphericalQG_Ham}, these are specific combinations of the vertex and edge labels, whereas $\Phi_0$ also contains the contribution from (scalar) matter Hamiltonian constraint. In the current paper, however, we shall regard $\Phi_\sigma$ and $R$ as some undetermined functions that are to be restricted by the requirement for the constraint algebra to be anomaly free. 
\subsubsection{Diffeomorphism constraint operator}\label{Sec:Diffeo_Def}
In this section we shall {\em define} the diffeomorphism constraint operator as the commutator of two Hamiltionian constraints
\be\label{DDef}
D[...]\propto \left[\hat H[N], \hat H[M]\right] = \sum\limits_{v,u}{N_v M_u \left[\left(\hat H_C+\hat H_R +\hat H_L\right)^v,\left(\hat H_C+\hat H_R +\hat H_L\right)^u\right]}.
\ee
Here the smearing function of the diffeomorphism constraint may involve canonical variables. It is easy to see that commutator is zero for $v=u$ because of symmetry. Also, the terms with $|v-u|\ge 2$ vanish, as the shifting part of one of the commutted operators does not affect the labels on the support of the multiplication part of the other operator. Therefore, the non-trivial contributions come only from the next-neighbour commutators for $u=v\pm 1$. Below we explicitly compute such commutators.

Consider for now $u=v+1$ (the $u=v-1$ case is very similar).We start by noting that 
\[
\left[\hat H_C^v, \hat H_C^{v+1}\right]=\left[\hat H_C^v, \hat H_R^{v+1}\right]=\left[\hat H_L^v, \hat H_R^{v+1}\right]=\left[\hat H_L^v, \hat H_C^{v+1}\right]=0.
\]
There are five non-zero commutators. Let us compute one explicitly
\bq\label{Comm1}
\left[\hat H_C^v, \hat H_L^{v+1}\right]&=&\left[\sum\limits_{\sigma=0,\pm 1}{\mbox{
\begin{picture}(100,15)(0,0)
 \put(0,5){\line(1,0){100}}
 \put(0,5){\circle*{5}}
 \put(50,5){\circle*{5}}
 \put(100,5){\circle*{5}}
 \put(50,-3){\makebox(0,0){${\bf v}$}}
 \put(50,12){\makebox(0,0){$+2\sigma\delta$}}
\end{picture}
}\Phi^v_\sigma}, \sum\limits_{\sigma_{1,2,3}=\pm 1}{\!\!\!\! \sigma_2 \sigma_3\mbox{
\begin{picture}(100,15)(0,0)
 \put(0,5){\line(1,0){100}}
 \put(0,5){\circle*{5}}
 \put(50,5){\circle*{5}}
 \put(100,5){\circle*{5}}
 \put(50,-3){\makebox(0,0){${\bf v}$}}
 \put(45,15){\makebox(0,0){$+\sigma_3\tfrac{\delta}{2}$}}
 \put(95,15){\makebox(0,0){$+\sigma_1\tfrac{\delta}{2}$}}
 \put(75,-3){\makebox(0,0){$+2\sigma_2$}}
\end{picture}
}\,\,\,\,R^{v+1}}\right]\nonumber\\
\nonumber\\
&=&\label{Comm1}\sum\limits_\sigma\!\sum\limits_{\sigma_{1,2,3}}\mbox{
\begin{picture}(150,15)(0,0)
 \put(0,5){\line(1,0){150}}
 \put(0,5){\circle*{5}}
 \put(75,5){\circle*{5}}
 \put(150,5){\circle*{5}}
 \put(75,-3){\makebox(0,0){${\bf v}$}}
 \put(70,15){\makebox(0,0){$+\delta\!\left(2\sigma+\tfrac{\sigma_3}{2}\right)$}}
 \put(145,15){\makebox(0,0){$+\sigma_1\tfrac{\delta}{2}$}}
 \put(110,-3){\makebox(0,0){$+2\sigma_2$}}
\end{picture}}\quad\sigma_2 \sigma_3\\
\nonumber\\
&&\times R^{v+1}\left[\Phi^{v}_\sigma\left(\mu_v+\sigma_3\tfrac{\delta}{2},k_v+2\sigma_2\right)-\Phi^{v}_\sigma \right]\nonumber
,
\eq
where only the affected (shifted) arguments of $R^{v+1}$ and $\Phi^v_\sigma$ have been written out explicitly. Similarly the other four non-trivial commutators of the adjacent vertex constraint operators are computed as follows

\bq\label{Comm2}
\left[\hat H_L^v, \hat H_L^{v+1}\right]&=&\sum\limits_{\sigma_{1,2,3}}\sum\limits_{\sigma_{4,5,6}}
\quad\mbox{
\begin{picture}(150,15)(0,0)
 \put(0,5){\line(1,0){150}}
 \put(0,5){\circle*{5}}
 \put(75,5){\circle*{5}}
 \put(150,5){\circle*{5}}
 \put(-5,15){\makebox(0,0){$+\sigma_3\tfrac{\delta}{2}$}}
  \put(35,-3){\makebox(0,0){$+2\sigma_2$}}
 \put(75,-3){\makebox(0,0){${\bf v}$}}
 \put(70,15){\makebox(0,0){$+\tfrac{\delta}{2}\!\left(\sigma_1+\sigma_6\right)$}}
 \put(145,15){\makebox(0,0){$+\sigma_4\tfrac{\delta}{2}$}}
 \put(110,-3){\makebox(0,0){$+2\sigma_5$}}
\end{picture}} \quad\sigma_2 \sigma_3\sigma_5\sigma_6\\
&&\times R^{v+1}\left[R^{v}\left(\mu_v+\sigma_6\tfrac{\delta}{2},k_v+2\sigma_5\right)-R^{v}\right],\nonumber
\\
\nonumber\\
\label{Comm3}
\left[\hat H_R^v, \hat H_L^{v+1}\right]&=&\sum\limits_{\sigma_{1,2,3}}\sum\limits_{\sigma_{4,5,6}}
\quad \mbox{
\begin{picture}(150,15)(0,0)
 \put(0,5){\line(1,0){150}}
 \put(0,5){\circle*{5}}
 \put(75,5){\circle*{5}}
 \put(150,5){\circle*{5}}
 \put(75,-3){\makebox(0,0){${\bf v}$}}
 \put(70,15){\makebox(0,0){$+\tfrac{\delta}{2}\!\left(\sigma_1+\sigma_6\right)$}}
 \put(145,15){\makebox(0,0){$+\tfrac{\delta}{2}\!\left(\sigma_3+\sigma_4\right)$}}
 \put(110,-3){\makebox(0,0){$+2\sigma_2$}}
\end{picture}}\quad\sigma_2 \sigma_3\sigma_5\sigma_6\\
&&\times\left[R^{v+1}R^{v}\left(\mu_v+\sigma_6\tfrac{\delta}{2},k_v+2\sigma_5\right)-R^{v} R^{v+1}\left(\mu_{v+1}+\sigma_3\tfrac{\delta}{2}, k_v+2\sigma_2\right)\right]\nonumber
,\\
\nonumber\\
\label{Comm4}
\left[\hat H_R^v, \hat H_C^{v+1}\right]&=&\sum\limits_\sigma\!\sum\limits_{\sigma_{1,2,3}}
\quad\mbox{
\begin{picture}(150,15)(0,0)
 \put(0,5){\line(1,0){150}}
 \put(0,5){\circle*{5}}
 \put(75,5){\circle*{5}}
 \put(150,5){\circle*{5}}
 \put(75,-3){\makebox(0,0){${\bf v}$}}
 \put(70,15){\makebox(0,0){$+\delta\!\left(2\sigma+\tfrac{\sigma_1}{2}\right)$}}
 \put(145,15){\makebox(0,0){$+\sigma_3\tfrac{\delta}{2}$}}
 \put(110,-3){\makebox(0,0){$+2\sigma_2$}}
\end{picture}}\quad \sigma_2 \sigma_3 \\
&&\times R^{v}\left[\Phi^{v+1}_\sigma - \Phi^{v+1}_\sigma\left(\mu_{v+1}+\sigma_3\tfrac{\delta}{2}, k_v+2\sigma_2\right)\right]\nonumber
,\\
\nonumber\\
\label{Comm5}
\left[\hat H_R^v, \hat H_R^{v+1}\right]&=&\sum\limits_{\sigma_{1,2,3}}\sum\limits_{\sigma_{4,5,6}}
\quad\mbox{
\begin{picture}(150,15)(0,0)
 \put(0,5){\line(1,0){150}}
 \put(0,5){\circle*{5}}
 \put(75,5){\circle*{5}}
 \put(150,5){\circle*{5}}
 \put(-5,15){\makebox(0,0){$+\sigma_1\tfrac{\delta}{2}$}}
  \put(35,-3){\makebox(0,0){$+2\sigma_2$}}
 \put(0,-3){\makebox(0,0){${\bf v}$}}
 \put(70,15){\makebox(0,0){$+\tfrac{\delta}{2}\!\left(\sigma_3+\sigma_4\right)$}}
 \put(145,15){\makebox(0,0){$+\sigma_6\tfrac{\delta}{2}$}}
 \put(110,-3){\makebox(0,0){$+2\sigma_5$}}
\end{picture}}\quad\sigma_2 \sigma_3\sigma_5\sigma_6 \\
&&\times R^{v}\left[R^{v+1}-R^{v+1}\left(k_v+2\sigma_2,\mu_{v+1}+\sigma_3\tfrac{\delta}{2}\right)\right].\nonumber
\eq

%
%
Note that the commutators in Eqs. (\ref{Comm1})-(\ref{Comm5}) all have the prefactor $N_v M_{v+1}$. As was pointed out earlier, there are also contributions with the prefactor $N_v M_{v-1}$, which can be brought to a very similar form after shifting $v \rightarrow v+1$ (and flipping the overall sign) in the summation over $v$. Then the overall coefficient would be $N_v M_{v+1} - N_{v+1} M_{v} = N \Delta M - M \Delta N$ (where $\Delta N \equiv N_{v+1}-N_v$), which resembles the standard classical expression for the shift vector in the diffeomorphism constraint, $N \partial {}^x M - M\partial {}^x N$.

At first glance, the obtained shift vector $L=N \partial {}^x M - M\partial {}^x N$ constitutes a restricted class of smearing vectors. For example, if $\partial M = 0$, we can\rq{}t have a constant $L=M\partial N$ for periodic boundary conditions on $N$. In this sense, the above construction does not contain all possible shifts. However, since $n$N and $M$ always appear in the form $N \partial {}^x M - M\partial {}^x N$, we can {\em extend} this definition by replacing the combination above with an arbitrary $L$.

Another potential issue with the definition (\ref{DDef}) is the dependence of the shift vector on the inverse metric components. While classically the metric exists and is invertible, it is not necessarily the case at the quantum level. 
Nevertheless, for the purposes of this paper one can replace the multiplication parts on the righthand side of Eqs. (\ref{Comm1})-(\ref{Comm5}) with arbitrary functions of the lattice labels, as will become clear in Section \ref{Sec:Closure}.

Comparing the commutators above, we notice that the shifting parts on the righthand side of Eqs. (\ref{Comm1}), (\ref{Comm3}), and (\ref{Comm4}) have similar form, which is the same as the form of $\hat H^v_R$. On the other hand, Eqs. (\ref{Comm2}) and (\ref{Comm5}) differ from the other three commutators and $\hat H^v_R$. Their shifting parts affect three vertices and two edges between them. Moreover, even though the structure of these two shifts appears similar the affected vertices and edges are distinct. 

Our ultimate goal is to investigate whether the complete system of constraints (Hamiltonian plus Diffeomorphism) can be closed under the commutator operation. In particular, this requires the Hamiltonian and Diffeomorphism constraints to have a similar structure. Thus the {\em spreading} of the shifting part mentioned above may make the closure of the constraints problematic. We now have two options: consider the conditions under which the {\em spread} terms cancel already in the Diffeomorphism constraint; or to keep going and only require closure at the stage when $[D[...], H[...]]$  is computed. For now we choose the latter and will revisit this discussion in Section \ref{Sec:Discussion}.

\subsubsection{Classical limit}\label{Sec:Classical_Limit}
Before we analyze closure of the constraint algebra, it is helpful to check the classical limit of the commutators obtained in the previous section and see if we recover the classical expression in Eq. (\ref{ClassDiff}) (up to the redefinition of the shift vector discussed at the end of Section \ref{Sec:Diffeo_Def}). We already saw that the resulting commutator constitutes a sum over vertices with a smearing coefficient $N \Delta M - M \Delta N$. Inside the sum there are five terms listed in Eqs. (\ref{Comm1})-(\ref{Comm5}). We shall now focus on their behaviour when $\delta, \epsilon \rightarrow 0$.

We first recall that in connection representation, the shifting part of an operator (that contains extrinsic curvature components) corresponds to an exponential. For instance, 
the angular and edge holonomies should be replaced with
\bq\label{CL_Ang}
\mbox{
\begin{picture}(100,15)(0,0)
 \put(0,5){\line(1,0){100}}
 \put(0,5){\circle*{5}}
 \put(50,5){\circle*{5}}
 \put(100,5){\circle*{5}}
 \put(50,-3){\makebox(0,0){${\bf v}$}}
 \put(50,12){\makebox(0,0){$(+\delta)$}}
\end{picture}
}&\rightarrow&  \exp\left(i\delta \gamma K_{\vp}^v\right) \quad \rm{and} \\
\label{CL_Rad}
\mbox{
\begin{picture}(100,15)(0,0)
 \put(0,5){\line(1,0){100}}
 \put(0,5){\circle*{5}}
 \put(50,5){\circle*{5}}
 \put(100,5){\circle*{5}}
 \put(50,-3){\makebox(0,0){${\bf v}$}}
 \put(75,-3){\makebox(0,0){$(+2)$}}
\end{picture}
}
 &\rightarrow& \exp\left(i\smallint_{v}^{v_+}\gamma K_x \md x\right)
\eq
respectively. The integral in Eq. (\ref{CL_Rad}) is taken over an edge of length $\epsilon$. Hence in the classical limit, this integral will be approximated by $\epsilon\gamma K_x^v$.\\

The (triad dependent) multiplication part of the operators has the form of a (infinitesimal) change in a label dependent function, which can be Taylor expanded to give an explicit infinitesimal factor of $\delta$ or $\epsilon$. For instance, in (\ref{Comm1}) we have 
\be\label{DeltaPhi}
\Delta \Phi^v_\sigma \equiv \Phi^{v}_\sigma\left(\mu_v+\sigma_3\tfrac{\delta}{2},k_v+2\sigma_2\right)-\Phi^{v}_\sigma \left(\mu_v, k_v\right) = A_\sigma^v\sigma_3\delta + B_\sigma^v\sigma_2\epsilon + O(\delta^2, \delta\epsilon, \epsilon^2),
\ee
with the coefficients $A_\sigma^v$ and $B_\sigma^v$ given by the corresponding $\mu_v$- and $k_v$-derivatives of $\Phi_\sigma^v$. Note that the multiplication factor in Eq.(\ref{Comm3}) is of similar form. Indeed, by adding and subtracting $R^v R^{v+1}$ it can be recast as $R^{v+1} \Delta R^v - R^v \Delta R^{v+1}$.

We now have all the ingredients to consider the classical limit of Eqs. (\ref{Comm1})-(\ref{Comm5}). In the first commutator, Eq. (\ref{Comm1}), we replace the holonomies according to Eqs. (\ref{CL_Ang}) and (\ref{CL_Rad}) along with the triad coefficient in (\ref{DeltaPhi}) to obtain
\bq
&\sum\limits_\sigma\!\sum\limits_{\sigma_{1,2,3}}\mbox{
\begin{picture}(150,15)(0,0)
 \put(0,5){\line(1,0){150}}
 \put(0,5){\circle*{5}}
 \put(75,5){\circle*{5}}
 \put(150,5){\circle*{5}}
 \put(75,-3){\makebox(0,0){${\bf v}$}}
 \put(70,15){\makebox(0,0){$+\delta\!\left(2\sigma+\tfrac{\sigma_3}{2}\right)$}}
 \put(145,15){\makebox(0,0){$+\sigma_1\tfrac{\delta}{2}$}}
 \put(110,-3){\makebox(0,0){$+2\sigma_2$}}
\end{picture}}\quad\sigma_2 \sigma_3 R^{v+1}\left[\Phi^{v}_\sigma\left(\mu_v+\sigma_3\tfrac{\delta}{2},k_v+2\sigma_2\right)-\Phi^{v}_\sigma \right]\nonumber\\
&=\sum\limits_\sigma\!\sum\limits_{\sigma_{1,2,3}}\sigma_2\sigma_3\exp\left(2i\sigma\delta \gamma  K_{\vp}^v\right)\exp\left(i \sigma_3\case{\delta}{2} \gamma  K_{\vp}^v\right)\exp\left(i\sigma_2\epsilon \gamma  K_{x}^v\right)\exp\left(i\sigma_1  \case{\delta}{2} \gamma K_{\vp}^{v+1}\right)\Delta\Phi_\sigma^v.\label{SC_1A}
\eq
In this factorized form, the summation over different $\sigma$\rq{}s can be performed separately. Recall that $\sigma$ takes values 0 or $\pm 1$, whereas all other $\sigma$\rq{}s can only be $\pm1$. Note also, that $\Phi_+=\Phi_-\neq \Phi_0$. Therefore, the summation over $\sigma$ contribites a factor of
\[
2\cos(2\delta \gamma K_{\vp}^v)\Delta\Phi_+ + \Delta \Phi_0\underset{\delta\to 0}{\longrightarrow}2\Delta\Phi_+ + \Delta \Phi_0 = A_1\sigma_3\delta + B_1\sigma_2\epsilon + O(\delta^2, \delta\epsilon, \epsilon^2),
\]
where $A_1$ and $B_1$ combine $A_\sigma^v$ and $B_\sigma^v$ respectively. Summation over $\sigma_1$ in (\ref{SC_1A}) contributes a factor of $2\cos(\case{\delta}{2}\gamma K_\vp^v)\underset{\delta\to 0}{\longrightarrow}2$. The remaining summation over $\sigma_2$ and $\sigma_3$ involves factors of $\sigma$\rq{}s inside the sums. Since $\sigma_2^2=\sigma_3^2=1$, we obtain
\bq
&&\sum\limits_{\sigma_{2,3}}\exp\left(i \sigma_3\case{\delta}{2} \gamma  K_{\vp}^v\right)\exp\left(i\sigma_2\epsilon \gamma  K_{x}^v\right)\sigma_2\sigma_3\left(A_1\sigma_3\delta + B_1\sigma_2\epsilon\right)\nonumber\\
&&=4iA_1\delta\cos\left(\case{\delta}{2} \gamma  K_{\vp}^v\right)\sin\left(\epsilon \gamma  K_{x}^v\right)+4iB_1\epsilon \sin\left(\case{\delta}{2} \gamma  K_{\vp}^v\right)\cos\left(\epsilon \gamma  K_{x}^v\right)\nonumber\\
&&=4i\delta\epsilon \gamma\left(A_1 K_x^v+\case{1}{2}B_1K_\vp^v \right)+o(\delta^2, \delta\epsilon, \epsilon^2).\label{Comm1B}
\eq
When considering the classical limit of (\ref{Comm2}), we can similarly expand the multiplication part as
\be\label{DeltaR}\nonumber
\Delta R^v \equiv  R^{v}\left(\mu_v+\sigma_6\tfrac{\delta}{2},k_v+2\sigma_5\right)-R^{v} \left(\mu_{v}, k_v\right)= A_2\sigma_6\delta + B_2\sigma_5\epsilon + O(\delta^2, \delta\epsilon, \epsilon^2)
\ee
 and perform the summation much like we did with (\ref{SC_1A}). Note that if an explicit factor of $\sigma_i$ is present inside the sum, the corresponding summation yields a sine function, which gives an infinitesimal contribution proportional to either $\delta$ or $\epsilon$. If such a factor is absent, the summation contributes a factor of $\cos(...)\underset{\delta\to 0}{\longrightarrow} 1$. Hence the number of $\sigma$\rq{}s determines the perturbative order of the commutator\rq{}s classical limit. Noticing that the extra factors of either $\sigma_5$ or $\sigma_6$ from the expansion of $\Delta R^v$ cancel one of the corresponding original $\sigma$\rq{}s in the summation, we conlude that (\ref{Comm2}) yields a higher (fourth) order contribution in the classical limit. The same conclusion applies to the classical limit of (\ref{Comm3}) and (\ref{Comm5}), but the contribution from (\ref{Comm4}) is of the same (second) order as that from (\ref{Comm1}). Expanding the difference term of (\ref{Comm4})
\[
-\Delta \Phi^{v+1}_\sigma \equiv \Phi^{v+1}_\sigma\left(\mu_{v+1}, k_v\right)-\Phi^{v+1}_\sigma \left(\mu_{v+1}+\sigma_3\tfrac{\delta}{2},k_v+2\sigma_2\right) = -A_\sigma^{v+1}\sigma_3\delta - B_\sigma^{v}\sigma_2\epsilon + O(\delta^2, \delta\epsilon, \epsilon^2),
\] 
we note that the coefficients $A_\sigma^{v+1}$ and $B_\sigma^v$ are proportional to the $\mu_{v+1}$- and $k_v$-derivatives of $\Phi^{v+1}_\sigma$. Repeating the steps we did for (\ref{SC_1A}) yields
\[
{\rm (\ref{Comm4})} \underset{\delta\to 0}{\longrightarrow} -4i\delta\epsilon \gamma\left(A_1^{v+1} K_x^v+\case{1}{2}B_1^v K_\vp^{v+1} \right)+o(\delta^2, \delta\epsilon, \epsilon^2).
\]
It is then clear that combining with (\ref{Comm1B}) results, up to a triad dependent coefficient, in
\[
{\rm (\ref{Comm1})+(\ref{Comm4})} \underset{\delta\to 0}{\longrightarrow}(A_1^{v+1}-A_1^v) K_x^v + B_1^v \left(K_\vp^{v+1}-K_\vp^v\right).
\]
Since the latter term constritutes the discrete version of $K_\vp^\prime$, and the commutators (\ref{Comm1}) and (\ref{Comm4}) are the leading terms in the overall $\left[\hat H[N], \hat H[M]\right]$ commutator, it is evident that an appropriate choice of $R$ and $\Phi_\sigma$ will yield the correct classical limit of the diffeomorphism constraint (\ref{ClassDiff}).
\section{Closure of the constraint algebra}\label{Sec:Closure}
In this section we shall investigate whether the diffeomorphism constraint operator defined in Section \ref{Sec:Diffeo_Def} forms a closed algebra with the Hamiltonian constraint operator given by (\ref{Ham_Def}). As was mentioned earlier in the Introduction, for closure it is necessary and sufficient that the commutator $\left[D[...], H[...]\right]$ \lq\lq{}resembles\rq\rq{} a Hamiltonian constraint with some lapse function $L$. Motivated by the classical expression, the this lapse function may in general be dependent on the triad operators. Moreover, the shift vector appearing inside the diffeomorphism constraint has an explicit metric dependence. In section \ref{Sec:Diffeo_Def}, we discussed how this issue can be overcome by using a general shift vector. It is easy to show, however, that such a replacement would not change the main conclusions. While the inverse metric in the shift of the diffeomorphism operator does indeed contribute to the commutator with a Hamiltonian constraint operator, the structure of the resulting operator (the shifting part) is unaffected by this extra term.

So far we have kept the triad dependence of the Hamiltonian constraint operator, i.e. the functions $\Phi_\sigma^v$ and $R^v$, unrestricted. These functions, however, affect neither the shifting part of the Hamiltonian constraint nor the structure of the shifting part of the commutators in Eqs. (\ref{Comm1})-(\ref{Comm5}). Therefore the shifting part of $\left[D[...], H[...]\right]$ must coincide with that of $H[...]$ itself, where each term affects at most two neighboring vertices and the edge between them. We can use the freedom in the triad dependence of $\Phi_\sigma^v$ and $R^v$ to require that all the terms in $\left[D[...], H[...]\right]$ that affect more labels vanish. 

There are quite a few combinations present in the expression for $\left[D[...], H[...]\right]$, but we start by considering two typical representatives: $\left[\widehat{{\rm Eq. (\ref{Comm3})}}, \hat H^{v+1}_L\right]$ and $\left[\widehat{{\rm Eq. (\ref{Comm2})}}, \hat H^{v+2}_L\right]$. The former has much overlap in the shifting parts of the two operators inside the commutator, which results in a compact overall shifting part. The latter, on the contrary, has little overlap in shifting parts, which yields a spread resulting shifting part. The explicit expressions for the compact commutator is given by
\bq
\label{Comm3A}
\left[\left[\hat H_R^v, \hat H_L^{v+1}\right],\hat H^{v+1}_L\right]&=&\sum\limits_{\sigma_{1,2,3}}\sum\limits_{\sigma_{4,5,6}}\sum\limits_{\sigma_{7,8,9}}
\mbox{
\begin{picture}(200,15)(0,0)
 \put(0,5){\line(1,0){200}}
 \put(0,5){\circle*{5}}
 \put(100,5){\circle*{5}}
 \put(200,5){\circle*{5}}
 \put(100,-3){\makebox(0,0){${\bf v}$}}
 \put(90,15){\makebox(0,0){$+\tfrac{\delta}{2}\!\left(\sigma_1+\sigma_6+\sigma_9\right)$}}
 \put(190,15){\makebox(0,0){$+\tfrac{\delta}{2}\!\left(\sigma_3+\sigma_4+\sigma_7\right)$}}
 \put(153,-5){\makebox(0,0){$+2(\sigma_2+\sigma_5+\sigma_8)$}}
\end{picture}}\quad\sigma_2 \sigma_3\sigma_5\sigma_6\sigma_8\sigma_9\nonumber\\
&&\times\left[R^{v+1}P^{v}\left(\mu_v+\sigma_9\tfrac{\delta}{2},k_v+2\sigma_8, \mu_{v+1}+\sigma_7\tfrac{\delta}{2}\right)\right. \\
&&\left.\quad-P^{v} R^{v+1}\left(\mu_{v+1}+(\sigma_3+\sigma_4)\tfrac{\delta}{2}, k_v+2(\sigma_2+\sigma_5)\right)\right],\nonumber
\eq
where
\[
P^{v}\left(k_{v-1},\mu_v,k_v, \mu_{v+1},k_{v+1}\right)=R^{v+1}R^{v}\left(\mu_v+\sigma_6\tfrac{\delta}{2},k_v+2\sigma_5\right)-R^{v} R^{v+1}\left(\mu_{v+1}+\sigma_3\tfrac{\delta}{2}, k_v+2\sigma_2\right).
\]
The second, non-compact, commutator is given by
\bq
\label{Comm3B}
\left[\left[\hat H_L^v, \hat H_L^{v+1}\right],\hat H^{v+2}_L\right]&=&\sum\limits_{\sigma_{1,2,3}}\sum\limits_{\sigma_{4,5,6}}\sum\limits_{\sigma_{7,8,9}}
\quad \mbox{
\begin{picture}(225,15)(0,0)
 \put(0,5){\line(1,0){225}}
 \put(0,5){\circle*{5}}
 \put(75,5){\circle*{5}}
 \put(150,5){\circle*{5}}
 \put(225,5){\circle*{5}}
 \put(75,-3){\makebox(0,0){${\bf v}$}}
 \put(0,15){\makebox(0,0){$+\sigma_3\tfrac{\delta}{2}$}}
 \put(70,15){\makebox(0,0){$+\tfrac{\delta}{2}\!\left(\sigma_1+\sigma_6\right)$}}
 \put(145,15){\makebox(0,0){$+\tfrac{\delta}{2}\!\left(\sigma_4+\sigma_9\right)$}}
  \put(225,15){\makebox(0,0){$+\sigma_7\tfrac{\delta}{2}$}}
 \put(40,-3){\makebox(0,0){$+2\sigma_2$}}
 \put(110,-3){\makebox(0,0){$+2\sigma_5$}}
 \put(185,-3){\makebox(0,0){$+2\sigma_8$}}
\end{picture}}\nonumber\\
&\times&\sigma_2 \sigma_3\sigma_5\sigma_6\sigma_8\sigma_9 R^{v+2} \\
&\times&\left[R^{v+1}\left(\mu_{v+1}+\sigma_9\tfrac{\delta}{2},k_{v+1}+2\sigma_8\right)-R^{v+1}\right]\left[R^{v}\left(\mu_{v}+\sigma_6\tfrac{\delta}{2}, k_v+2\sigma_5\right)-R^{v}\right].\nonumber
\eq
Looking at the two equations above, one can notice that a more compact shifting part is accompanied by a more sophisticated triad dependent factor and vice versa, the shifting part in Eq. (\ref{Comm3B}) is non-compact, but the multiplication part is fairly simple. Importantly, the terms corresponding to different vertices or different $\sigma$\rq{}s appear with different  combinations of lapse functions, hence must vanish independently. If we now require that all the non-compact terms equal zero, including the term in Eq. (\ref{Comm3B}), this would result in only trivial solutions for $R^v$ that are merely constant and do not depend on the triad components. This appears too restrictive to allow for physically interesting models.
 
\section{Discussion}\label{Sec:Discussion}
We investigated a possibility of an anomaly free lattice-based quantization motivated by loop quantum gravity ideas. We primarily focused on a spherically symmetric case, but some of the conlusions apply to more general lattices, including the cubic one. The major requirement that we imposed on the quantum constraint operators was their off-shell closure, i.e. closure of the commutator algebra of the constraints without any reference to specific states. 

The idea of off-shell closure can be used in two different ways. On the one hand, it may be used in a {\em constructive} manner: starting with a single constraint operator (Hamiltonian) one can {\em define} the Diffeomorphism constraint operator as commutator of two Hamiltonians, as is done in Section \ref{Sec:Diffeo_Def}. If the newly defined operator weakly commutes with another Hamiltonian {\em and} both the Diffeomorphism constraint and the constraint algebra have the right classical limit, the system of constraints would constitute an adequate quantization of the hypersurface deformation algebra. So technically, one \lq\lq{}only\rq\rq{} needs to construct a suitable Hamiltonian constraint and the rest of the system will follow.

On the other hand, off-shell closure can be used to test whether certain proposed constructions of the Hamiltonian constraints can yield a consistent quantization. One would just follow the same procedure as explained in Sections \ref{Sec:Diffeo_Def} -- \ref{Sec:Closure}, including checking the classical limit.

In order for the constraint commutators to weakly vanish off-shell each commutator must be represented by a linear combination of the constraints. In the lattice framework, this is possible only if the action of the commutators on the lattice states has the same form as the action of the original constraints, i.e. the original constraints and commutators affect the same lattice labels. In this paper, we focused on the spherically symmetric case (Section \ref{sec:HamConst}) and specifically on the Hamiltonian constraint operator motivated by Ref. \cite{SphericalQG_Ham}. We kept some freedom in the triad dependent factors in the Hamiltonian, but it was still not possible to obtain an anomaly free constraint algebra. The main issue was the {\em spreading} of the shifting part of the constraints after computing commutators, which was incompatible with the form of the original Hamiltonian constraint and could not yield closure. This phenomenon appears quite generic for any lattice quantization of this type. For instance, an analogous quantization based on a cubic lattice (motivated by Ref. \cite{QuantCorrPert}) would exhibit similar spreading - although in a more sophisticated way - owing to the three-dimensional nature of the lattice.

We would also like to remark on the relation of the construction considered in the current paper to the anomaly free results obtained in a series of papers utilizing effective perurbative approach \cite{ConstraintAlgebra, ScalarGaugeInv} and \cite{ScalarHol} -- \cite{ScalarHolInv}. These papers present perturbatively consistent (at the effective level) systems of constraints that are first class and allow for gauge invariant formulation of equations of motion. As was shown in Section \ref{Sec:Classical_Limit}, the terms corresponding to the spread commutators are  subdominant in the classical limit. They are of higher order in the perturbative parameter than the main terms showing up in the effective formulation considered in the aforementioned papers. Interestingly, the spread (non-local) operators can still be mimicked at the effective level by including higher order spatial derivatives \cite{QuantCorrPert, HigherSpatDer}. As was pointed out in \cite{HigherSpatDer}, such inclusion leads to related anomaly issues. Another interesting approach to anomaly-free quantization in the spherically symmetric sector was considered in \cite{GP_Abel}. It involves rescaling of the lapse and shift prior to quantization, which leads to \lq\lq{}Abelianization\rq\rq{} of the Hamiltonian constraint. Such a quantization scheme is consistent and contains some discreteness effects after Abelianization, but it is not clear whether it allows a representation of hypersurface-deformation algebra with the correct classical limit.

In summary, the attempted lattice-based quantization of spherically symmetric hypersurface deformation constraint algebra has not been successful. It has, however, provided an insight as to what possible consistent construction could be. In order to have a closed constraint algebra, one needs to make sure that the structure (especially the shifting part) of the original Hamiltonian constraint operator has to be the same as that of the second order commutators of the Hamiltonians $\left[H[N_1],\left[H[N_2], H[N_3]\right]\right]$. Even if one starts with simple holonomies in $H[N]$, the commutators will have more complicated loop structure. This therefore suggests that one should already consider all possible holonomies in the original Hamiltonian, including loops of arbitrarily large size. At the moment, there is no clear understanding as to what a good parametrization of such loops could be, but it may still be possible to construct in the future.

\section*{Acknowledgements}
The author is grateful to Martin Bojowald for  discussions and valuable suggestions for improving the manuscript.

\end{document}